\documentstyle[aps,preprint]{revtex}
 
\def\lsim{\mathrel{\rlap{\lower4pt\hbox{\hskip1pt$\sim$}}
    \raise1pt\hbox{$<$}}}         %less than or approx. symbol
\def\gsim{\mathrel{\rlap{\lower4pt\hbox{\hskip1pt$\sim$}}
    \raise1pt\hbox{$>$}}}         %greater than or approx. symbol

\def\Pom{{\bf I\!P}}
 
\begin{document}
\draft
\title{How to measure the pion structure function
at HERA}
\author{H.Holtmann$^{1}$, G.Levman$^{2}$,
 N.N.Nikolaev$^{1,3}$, A.Szczurek$^{1,4}$,
 and J.Speth$^{1}$ }
\address{
$^{1}$Institut f\"ur Kernphysik, Forschungszentrum  J\"ulich GmbH.,\\
52425 J\"ulich, Germany \\
$^{2}$  Department of Physics, University of Toronto,\\
60 St. George Street, Toronto, Ontario M5S 1A7, Canada \\
$^{3}$  Landau Institute for Theoretical Physics, \\
 GSP-1, 117940, ul.Kosygina 2, 117334 Moscow, Russia \\
$^{4}$ Institute of Nuclear Physics,\\
ul. Radzikowskiego 152, PL-31-342 Krak\'ow, Poland}
%\date{\today}
\maketitle

\begin{abstract}
We suggest a method of determination of the pion structure function
down to $x\simeq 10^{-4}$ based on semi-exclusive deep inelastic
scattering off protons. The idea is to exploit the nonperturbative
$\pi N$ and $\pi\Delta$ Fock components of the nucleon, which
contribute significantly to deep inelastic scattering and dominate
the fragmentation of protons into fast neutrons and deltas. The
intrinsic factorization properties of the semi-exclusive cross section
give a good test for the validity of this approach.
\end{abstract}
\pacs{}

Up to now the only feasible method to extract the pion structure was
the $\pi N$ Drell-Yan production. The disadvantages of this method are
that the attainable luminosity is low and that only the valence part
of the pion structure function at rather large $x$ ($\gsim0.2$) can be
studied. An extension of our knowledge of the pion structure function
is possible by using virtual pions around the nucleon as targets in
deep inelastic scattering.  These pions arise naturally as a
consequence of the pion-nucleon coupling which leads to an admixture
of the $\pi N$ Fock state in the light-cone nucleon. Interaction of
high-energy projectiles (nucleons, pions, leptons) with the virtual
pion of the $\pi N$ Fock state of the proton is a typical stripping
reaction, in which the momentum distribution of the spectator nucleon
reflects the momentum distribution in the $\pi N$ (meson-baryon) Fock
state. For instance, the differential cross section of the inclusive
reaction $\pi p
\rightarrow nX$ takes the form (Fig.~1a)
\begin{equation}
{d\sigma(\pi p \rightarrow nX) \over dz dp_{\perp}^{2}}=
{2\over 3}f_{N/p}(z,p_{\perp}^{2})\sigma_{\pi \pi}(s_{\pi \pi}),
\label{sigmapion}
\end{equation}
where $f_{N/p}(z,p_{\perp})$ is the flux of nucleons with the
transverse momentum $p_{\perp}$ and the fraction $z$ of the
proton's light-cone momentum. The $\pi\pi$ total cross
section enters at the center of mass energy squared
$s_{\pi\pi}=(1-z)s_{\pi p}$ and the factor ${2\over 3}$ comes
from the isotopic Clebsch-Gordan coefficients.
 
The salient feature of Eq.~(\ref{sigmapion}) is that the flux
$f_{N/p}(z,p_{\perp})$ as well as its counterpart
$f_{\Delta/p}(z,p_{\perp})$ for the $p\rightarrow \Delta$
inclusive fragmentation, is a universal property of the $\pi N$,
$\pi \Delta$ Fock state, which does not depend on the projectile.
In the light-cone representation they read (see \cite{Z92,HSS93})
\begin{equation}
f_{N/p} (z,p_{\perp}^2) =
 \frac{3 g_{p \pi^0 p}^2}{16\pi^2}
 \frac{ \bigl[ (1-z)^2m_{N}^2+p_{\perp}^2 \bigr]}
 {z^2(1-z)(m_{N}^2 - M_{\pi N}^2)^2} |F(z,p_{\perp}^2)|^2 ,
\label{fnucleon}
\end{equation}
and
\begin{equation}
f_{\Delta/p} (z,p_{\perp}^2) =
 \frac{2 g_{p \pi^- \Delta^{++}}^2}{16\pi^2}
\frac{\bigl[ (zm_{N}+m_{\Delta})^2 + p_{\perp}^2 \bigr]^2
      \bigl[ (zm_{N}-m_{\Delta})^2 + p_{\perp}^2 \bigr]}
{ 6 z^4(1-z) m_{\Delta}^2 (m_{N}^2 - M_{\pi\Delta}^2 )^2 }
               |F(z,p_{\perp}^2)|^2,
\label{fdelta}
\end{equation}
These fluxes are nonperturbative quantities. Their absolute
normalization is fixed by the nonperturbative strong couplings known
from low-energy physics: $g^2_{p\pi^0 p}/4\pi=13.6$ \cite{TRS91} and
$g^2_{p\pi^-\Delta^{++}}/4\pi=12.3\>GeV^{-2}$ \cite{HHS89}. The vertex
light-cone form factors $F(z,p_{\perp}^2)$ are parameterized as
\cite{Z92,HSS93,HSS94}
\begin{equation}
F(z,p_{\perp}^2) =
\exp\left[- \frac{R_{MB}^2}{2}
  (M_{MB}^2(z,p_{\perp}^2) - m_N^2)\right]
\label{formfac}
\end{equation}
in terms of still another nonperturbative parameter---the radii
$R_{MB}$ of the meson-baryon Fock state. Here $M_{MB}(z,p_{\perp}^2)$ is
the invariant mass of the intermediate two-body meson-baryon Fock
state
\begin{equation}
M_{MB}^2(z,p_{\perp}^2) =
\frac{m_B^2 + p_{\perp}^2} {z} \; + \;
\frac{m_M^2 + p_{\perp}^2} {1-z} \; .
\end{equation}
The nonperturbative radius $R_{MB}$ is the only adjustable parameter
and it describes the $z$ and $p_{\perp}$ distribution of the observed
neutrons (deltas). The pion-exchange is a well defined
nonperturbative dynamical model for the $p\to n$, $\Delta$
fragmentation processes, and it has been shown
\cite{Z92,HSS94,AB85,ABK81,AG81} that the simple formulas
Eqs.~(\ref{fnucleon},\ref{fdelta}) give an excellent quantitative
description of the experimental data on high energy neutron and
$\Delta^{++}$ production in a broad range of $z$ (Near the kinematical
boundary $1-z \ll 1$ the fluxes Eqs.~(\ref{fnucleon},\ref{fdelta})
must be modified for the reggeization of pions \cite{BKP75}, but we
shall not consider this region of $z$). The pion-exchange contribution
exhausts the observed cross section at $z\sim 0.7$--$0.8$. This
agreement shows that the production of fast nucleons and deltas with
$z\sim 0.7$--$0.8$ is a small background to the pion-exchange
contribution.  The recent analysis \cite{HSS94} gave $R_{\pi N} =
0.93GeV^{-1}$ and $R_{\pi\Delta}=1.02GeV^{-1}$.  Notice, that once the
fluxes $f_{N/p}(z,p_{\perp})$ and $f_{\Delta/p}(z,p_{\perp})$ are
known, one can use the data on the $\pi p\rightarrow n(\Delta)X$
inclusive reactions for the determination of the high-energy pion-pion
total cross section (see
\cite{ZS84,H'76} and references therein).
 
Capturing on this remarkable success of the pion exchange in
hadronic reactions, we expect that the semi-inclusive reactions
\begin{equation}
a)\quad ep \rightarrow e'nX \qquad\qquad b)\quad ep\to e'\Delta X
\end{equation}
in the properly chosen kinematical domain also will be dominated by
the pion exchange (Sullivan \cite{S72}) mechanism of Fig.~1. If this is
the case, then the straightforward generalization of
Eq.~(\ref{sigmapion}) to semi-inclusive deep inelastic scattering is
\begin{equation}
\frac{d\sigma(ep\rightarrow e'n X)}
     {dx dQ^2 dz dp_{\perp}^2 } =
{2\over 3}f_{N/p}(z,p_{\perp}^2)
  K(x,Q^2) F_2^{e\pi}(x_{\pi},Q^2),
\label{4DCS}
\end{equation}
where $F_2^{e\pi}(x_{\pi},Q^2)$ is the structure function of the pion;
$x_{\pi}=x/(1-z)$ is the Bjorken variable in the electron-pion deep
inelastic scattering with the obvious kinematical restriction $0 < x <
1-z $, and $K(x,Q^2)$ is the standard kinematical factor
\begin{equation}
K(x,Q^2)={4\pi\alpha^2\over Q^4}
  {1\over x}\left[1-y+{y^2\over2}\right],\qquad y=\frac{Q^2}{xs},
\end{equation}
assuming for the sake of simplicity $2xF^{e\pi}_1(x)=F^{e\pi}_2(x)$.
Knowing all kinematical variables and trusting the theoretical
prediction for $f_{N/p}(z,p_{\perp}^2)$, one can invert
Eq.(\ref{4DCS}) and determine the pion structure function from the
experimentally measured semi-inclusive cross section. In the HERA
experiments, one can go down to the region of very small
$x_{\pi}\gsim 10^{-4}$. This will be an enormous expansion of the
studied kinematical region compared to the $\pi N$ Drell-Yan
experiments, which can not go much below $x_{\pi} \sim 0.1$
\cite{SMRS92}. Besides, such a determination of the pion structure
function at HERA will allow to study the scaling violations in the
pion structure function in a broad range of $(x_{\pi}, Q^{2})$,
which hardly is possible in the Drell-Yan experiments.  The aim of the
present communication is to analyze kinematical conditions under which
the pion structure function can be extracted.
 
From the purely experimental point of view, the semi-inclusive
reaction $ep\rightarrow e'nX$ is being studied already by the
ZEUS collaboration, which has installed a test forward neutron
calorimeter (FNC) to complement its leading proton spectrometer
\cite{ZEUS94}. This FNC was tested with neutrons from inclusive
proton-beam gas interactions, and an excellent agreement between
the measured spectra and the pion-exchange predictions was found.
 
From the theoretical point of view, the principal issue is the
relevance of the pion-exchange mechanism to deep inelastic scattering
off protons. The pion exchange diagrams of Fig.~1 correspond to deep
inelastic scattering off the nonperturbative pion-induced sea in
nucleus. There is ample experimental evidence for such a
nonperturbative sea in nucleons, which presently can be evaluated
parameter-free making use of the pion structure function as determined
from the $\pi N$ Drell-Yan data \cite{SMRS92} and the universal fluxes
Eqs.~(\ref{fnucleon},\ref{fdelta}).  Further evidence for the
importance of pion-induced (and $\rho$-induced) sea in DIS comes from
the observed Gottfried Sum Rule violation, connected to the $\bar
u$-$\bar d$ asymmetry \cite{Z92,HM90,SST91,HSB91,SS93,SSG93,SH93},
which can be calculated parameter-free and agrees with the
experimental data \cite{Z92,HSS94}.  The predicted $\bar u$-$\bar d$
asymmetry was confirmed \cite{HNSS94} by the measurement of the $\bar
u(x)/\bar d(x)$ ratio in the recent NA51 Drell-Yan experiment
\cite{K94}.  Furthermore, the nonperturbative meson-induced sea was
shown \cite{HNSS94} to dominate the nucleon sea at $x\gsim 0.1$.
 
Our principal task is to find the kinematical domain in which
the semi-inclusive reaction $ep\to e´nX$ is dominated by
the pion exchange contribution. The semi-inclusive production
of neutrons with $z\sim 0.8$ turns out to be the optimal
kinematical domain, and it also corresponds to the domain
in which the semi-inclusive cross section is largest.
Then, we suggest tests of the pion-exchange dominance and
evaluate some of background reactions which also lead to
production of fast neutrons.
 
Let us start with the charge-exchange reaction $p \rightarrow n$ and
let us consider first the $p_{\perp}$-integrated semi-inclusive cross
sections. The corresponding flux of neutrons $f_{n/p}(z)= {2\over
3}f_{N/p}(z)$ is shown in Fig.~2. The salient feature of the
pion-exchange mechanism is that the corresponding neutrons are fast
and the spectrum of neutrons has its maximum at $z\sim 0.6-0.8$.  This
important property can easily be understood: evidently, heavy nucleons
carry larger fraction of momentum $z$ than light pions. The expected
counting rates can be judged by the total number of virtual pions in
the nucleon: $n_\pi(\pi N)\approx 0.18$, $n_\pi(\pi\Delta)\approx
0.06$\cite{HSS94} , which shows that the deep inelastic scattering on
pions, accompanied by $p\to n$, $\Delta$ fragmentation, will have a
statistics only by one order in magnitude lower than for the $ep$
scattering.
 
The background to the pure pion exchange comes from interaction with
Fock states which contain heavier mesons $M=K,\rho,\omega...$
Evidently, in such states the heavy mesons $M$ will carry larger
fraction of the momentum of the $MN$ state, and the heavy meson
exchange will contribute to the spectrum of neutrons at smaller $z$ as
compared to the pion exchange (we do not discuss here the region of $1-z
\ll 1$ where the reggeization of mesons becomes important). In Fig.~2.
we show the effect of the $\rho N$ Fock state, which was found to be
important at small $z$ \cite{HSS94}. Because of many spin couplings
the formulas for the corresponding fluxes are too lengthy to be
reproduced here; they can be found elsewhere
\cite{HSS93}.  Evidently, zooming at $z \sim 0.7-0.8$ one can
eliminate much of the $\rho$-exchange background. A still better
separation of the $\pi$ and $\rho$ exchange can be achieved if one
compares the $p_{\perp}$ distributions for the two mechanisms.  With
the radii $R_{MN}$ as determined in \cite{HSS94}, the $\rho$ exchange
gives a broader $p_{\perp}$ distributions than the pion exchange
(Fig.~2). If one selects only events with $p_{\perp}^2 < 0.1 \,
(GeV/c)^2$, then, without much loss of statistics, one can
significantly enhance the relative contribution of the pion exchange,
which is demonstrated in Fig.~4.
 
We wish to emphasize that, as a matter of fact, the $\rho$-exchange
contribution is not a real background, because it is natural to
expect that structure functions of the pion and the $\rho$-meson
are close to each other. Because of the different spin structure,
the $\rho$  and $\pi$  contributions do not interfere, and the
total semi-inclusive cross section will be proportional to
\begin{equation}
f_{N/p}^{(\pi N)} (z)F_{2}^{\pi}(x_{\pi},Q^{2})+
f_{N/p}^{(\rho N)}(z)F_{2}^{\rho}(x_{\pi},Q^{2})
\approx
[f_{N/p}^{(\pi N)}(z)+f_{N/p}^{(\rho N)}(z)]F_{2}^{\pi}(x,Q^{2}) \, ,
\label{totflux}
\end{equation}
Therefore, Eq.~(\ref{4DCS}) will hold in a broader range of $z$, if
the flux of nucleons $f_{N/p}(z)$ in Eq.~(\ref{4DCS}) is generalized
to include the effect of $N\rho$ states as well. Still, we feel it is
safer to concentrate on $z\sim 0.7$--$0.8$, which is also the region
from where the dominant part of the semi-inclusive cross section is
coming.
 
Above we have considered the {\it elastic} fragmentation $p\rightarrow
n$. One can also consider the excitation of the proton into resonances
and continuum states, which then will decay into the observed
neutron. The prominent excitation channel is the production of
$\Delta$ states, which is also dominated by the same pion (and $\rho$)
exchange mechanism; we shall comment more on this reaction
below. Another type of background is the diffractive excitation of the
proton into the prominent $N^{*}$ resonances and/or the low mass
$N\pi$, $N\pi\pi,\dots$ continuum states, whose decays can produce
fast neutrons with large $z$.  Such diffractively produced states
carry $\approx 100\%$ of the protons momentum. The underlying
mechanism is the pomeron $(\Pom)$ exchange process of Fig.~1d, in
which the photon interacts with the pomeron. The relative fraction of
the pomeron-exchange and pion-exchange reactions must not change from
the $pp$ scattering to the deep inelastic scattering energy region. In
$pp$ scattering, the total cross section of diffractive excitation of
$N^{*}$ resonances and $N\pi,\,N\pi \pi$ continuum states is $\lsim
1\>mb$ \cite{ZZ88}, which is much smaller than the $p\to n$
charge-exchange cross section of $\sim 10\>mb$
\cite{AB85,ABK81,AG81,ZS84,H'76}. Thus, we conclude that the
background from the diffractive excitation of protons is negligibly
small. Furthermore, the pion and pomeron exchange reactions differ by
the rapidity distributions of secondary hadrons: in the pion-exchange
reaction the rapidity gap between the observed neutron and the
hadronic debris from the $\gamma^*\pi$ collision is small,
$\Delta\eta\sim \log(1/(1-z))\sim 1$, whereas in the diffractive
pomeron-exchange reaction there will be a large rapidity gap between
the decay products of $N^*$ and hadrons from the $\gamma^*\Pom$
interaction, which will allow to experimentally separate the two
reaction mechanisms.
 
In the suggested mechanism of the semi-inclusive neutron production,
the differential cross section Eq.~(\ref{4DCS}) is a product of the
universal flux factor which only depends on $z$ and the structure
function $F_{2}^{\pi}(x_{\pi},Q^{2})$ which is a function of $x_{\pi}=
x/(1-z)$. This factorization property allows an important cross check
of the model: Binning the semi-inclusive cross section data as a
function of $z$ at different fixed values of $x_{\pi}$ one can verify
that the shape of the flux factor as a function of $z$ does not depend
on $x_{\pi}$. Furthermore, this $z$-dependence must come out identical
to the $z$-dependence of the inclusive spectra of neutrons from the
hadronic $pN$ interactions. Remarkably, the FNC of the ZEUS
collaboration enables the latter cross check to be performed {\sl in
situ}, directly comparing the spectra of neutrons from inclusive
beam-gas interactions and from deep inelastic $ep$ scattering. Such a
comparision of the two spectra will allow to verify that the background
contribution to $z\sim 0.7$--$0.8$ from deep inelastic scattering off
the baryonic core is as small as in hadronic reactions. Reversing the
argument, one can determine the $x_{\pi}$ dependence of the pion
structure function changing $x$ at fixed values of $z$ and verify that
this $x_{\pi}$ dependence comes out the same at all values of $z$. Our
conclusion is that the separation of the pion exchange contribution
and determination of the pion structure function is experimentally
feasible, the possible backgrounds to the pion exchange are under
reasonably good control, and important {\sl in situ} cross checks of
the pion-exchange mechanism are possible.
 
The above discussion is fully applicable to the semi-inclusive
production of $\Delta^{++}$. The longitudinal momentum distribution of
$\Delta^{++}$ is shown in Fig.~2c. As in the $p \rightarrow n$ case,
the contributions from the $\pi$ and $\rho$ exchange mechanism are
fairly well separated with the $\pi$-exchange contribution dominating
large $z$.  Measuring the $\Delta^{++}$ production at HERA will
require good experimental resolution of both proton and $\pi^{+}$
resulting from the $\Delta^{++}$ decay, which requires multitrack
identification of the leading proton spectrometer.  The ZEUS
collaboration has such a device operating at HERA \cite{ZEUS93}.  The
measurements of the $\Delta^{++}$ production are important for the
direct evaluation of the contribution of the two-step process $p
\rightarrow \Delta \rightarrow n\pi$ to the spectrum of neutrons.
The $\Delta$ decay background to the spectrum of neutrons is
small. Isospin symmetry considerations  imply that the relative
contamination of the neutron spectra $n_\pi(\pi \Delta) \, / \, (3 n_\pi(\pi
N)) \approx 0.1$ is significantly smaller than that for the proton
spectra $(7n_\pi(\pi\Delta)) \, / \, (9n_\pi(\pi N)) \approx 0.3$ .

The dominance of the isovector exchange mechanism, both $\pi$ and
$\rho$ exchange, predicts very definite ratios of cross sections
for the $n/p$ and $\Delta^{++} / \Delta^{0}$ production:
$\sigma(p \rightarrow n)/\sigma(p \rightarrow p) = 2$ and
$\sigma(p \rightarrow \Delta^{++}) / \sigma(p\rightarrow \Delta^{0})
= 3$. The leading-proton spectrometer (LPS) installed at the ZEUS
detector allows measurement of the spectrum of protons in the
interesting region of $z\sim 0.6$--$0.8$. Here one must be aware
of the competitive pomeron-exchange contribution of Fig.~1d.
to the spectrum of leading protons. The salient feature of
pomeron-exchange is the approximately factorizable cross section
of excitations of large masses
\begin{equation}
\frac{1}{\sigma_{tot}(ap)}\frac{d\sigma(ap\to Xp)}{M^2\>dt}
\Biggr|_{t=0}\simeq
 \frac{A_{3\Pom}}{M^2+Q^2}
\end{equation}
Here $A_{3\Pom}\simeq 0.16\>GeV^{-2}$ \cite{C85} is the so-called
triple-pomeron constant which must be approximately the same in the
hadronic reactions, the real photoproductions and the deep inelastic
scattering \cite{NZ'92,NZ94}. The reaction of Fig.~1d can be
interpreted as a convolution of the flux of pomerons in the $p\Pom$
system with the structure function of the pomeron.  Both normalization
of the flux of pomerons in the proton and of the pomeron structure
function are convention dependent, because for the observable cross
section only the product of these two factors is needed.  The analysis
\cite{NZ'92,NZ94,BGNPZ94} has shown that the (anti)quark-gluon content
of the pomeron is similar to that of the $\pi^0$. If we choose the
pomeron structure function to have the same absolute normalization as
the pion structure function, then the observed flux of proton due to
pomeron exchange can be estimated as
\begin{equation}
f_{\Pom/N} (z) = {1 \over {1-z}}  \frac {A_{3\Pom}} {B_{3\Pom}}
\end{equation}
where $B_{3\Pom}$ is the diffraction dissociation slope
parameter. With $A_{3\Pom} \approx 0.16 (GeV/c)^{-2}$ and $B_{3\Pom}
\approx 6 (GeV/c)^{-2}$, as borrowed from the Regge phenomenology of
real photoproduction \cite{C85}, $A_{3\Pom}/B_{3\Pom} \approx 0.025$
\cite{NZ'92,NZ94}.  As clearly seen from the Fig.~2a rather good
separation of the pion and pomeron exchange mechanisms can be achieved
in the longitudinal momentum spectrum. The pion exchange mechanism is
dominant for $0.6 < z < 0.8$.

In conclusion, we have calculated the longitudinal and perpendicular
momentum distributions of leading protons, neutrons and $\Delta$'s in
high-energy electron-proton collisions with kinematics suitable for
HERA. We find that a large part of the phase space is populated
predominantly through the one-pion exchange mechanism, which is a
dominant and well-defined nonperturbative mechanism of $p\to n$ and
$p\to \Delta$ fragmentation. The background to the pion-exchange was
shown to be small.  This can be used to study the pion structure
function in a very small-$x$ region. The FNC of the ZEUS detector
allows to measure the leading neutrons and is most promising in this
respect. Tests of the factorization property of the semi-inclusive
cross section and the {\it in situ} comparison of the spectra of
neutrons from deep inelastic scattering and beam-gas interactions
allow independent checks of the pion-exchange mechanism.
 
{\it Acknowledgements.} This work was supported in part by the Polish
KBN grant 2 2409 9102 and by the INTAS grant 93-239.

%\bibliographystyle{my}
%\bibliography{dis} 

\begin{figure}
\caption{
The pion-exchange contribution to the inclusive production of neutrons
in (a) $\pi p$ scattering and (b) deep inelastic scattering and to the
inclusive production of $\Delta^{++}$ ($\Delta^0$) (c) in deep
inelastic scattering. Diagram (d) shows the diffractive production of
$N^*$'s by pomeron exchange.}
\end{figure}
 
\begin{figure}
\caption{
Longitudinal momentum distribution of (a) neutrons, (b) protons and
(c) $\Delta^{++}$. The contributions of the $\pi$ and $\rho$ exchange
mechanisms are shown by the dashed and dotted line, respectively.  The
contribution from the pomeron exchange mechanism to the $p\to p$
fragmentation is shown by the dash-dotted line.}
\end{figure}
 
\begin{figure}
\caption{
Inclusive perpendicular momentum distribution of (a) neutrons and (b)
$\Delta^{++}$.  The contributions corresponding to the $\pi$ and
$\rho$ exchanges are shown by the dashed and dotted line,
respectively.}
\end{figure}
 
\begin{figure}
\caption{
Longitudinal momentum distribution of neutrons (see Fig.~2a.) with the
extra condition $p_{\perp}^2 < 0.1 (GeV/c)^2$ (lower curves) compared
to the unconstrained one (upper curves, see Fig.~2a.). The $\pi$
exchange contributions are shown by dashed lines; the $\rho$ exchange
contributions by the dotted lines.}
\end{figure}
 
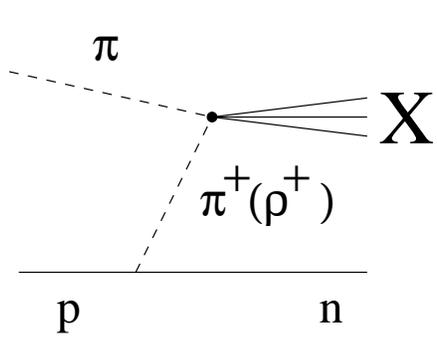
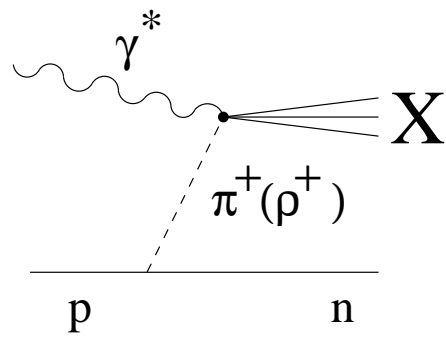
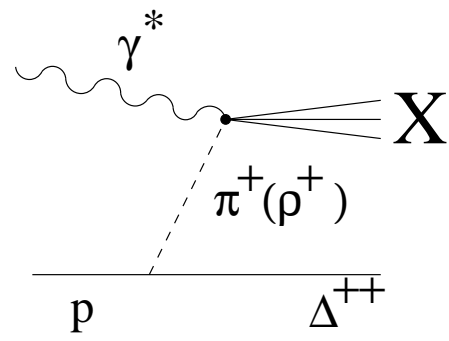
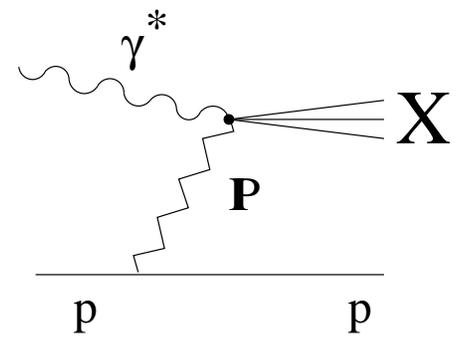
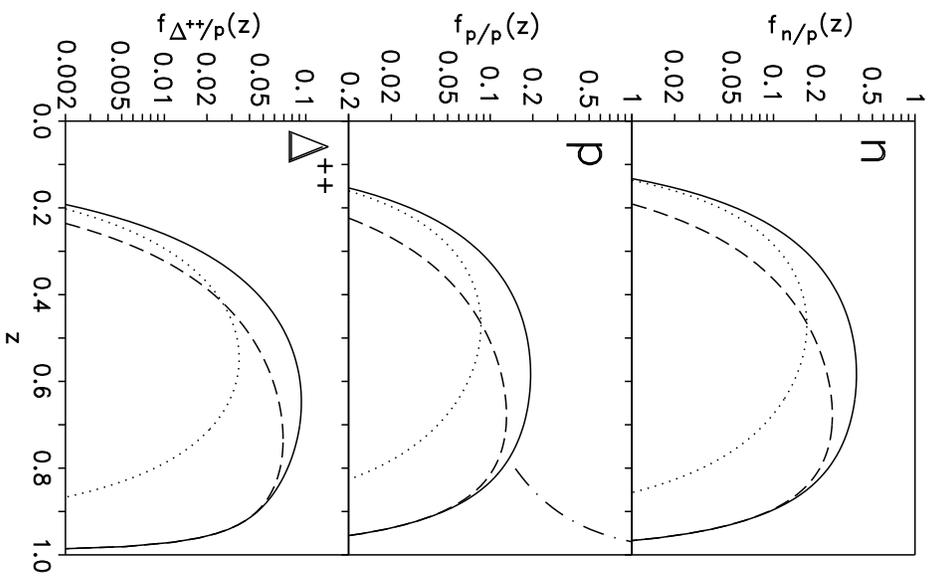
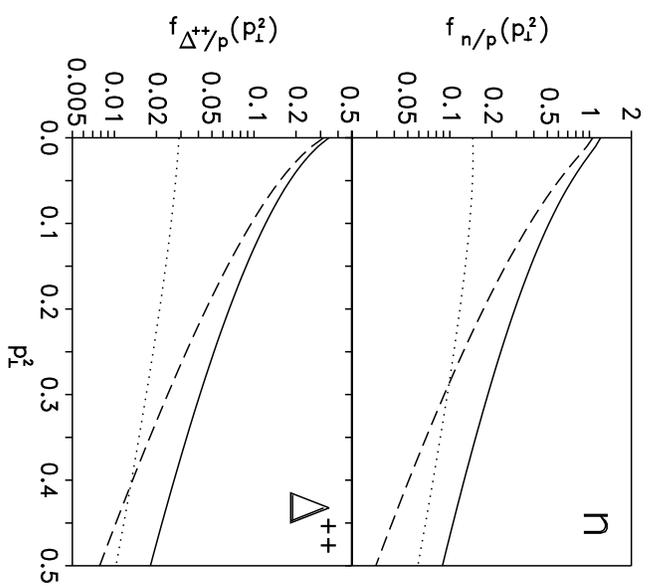
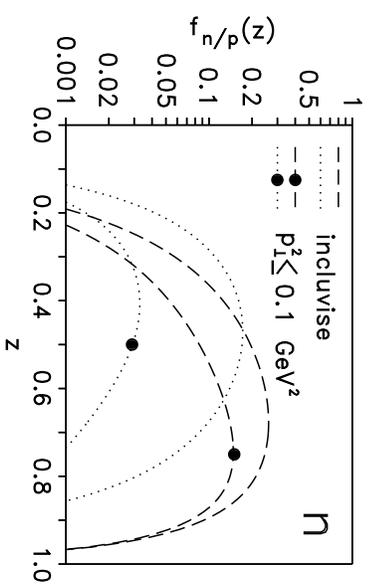
\end{document}